\documentclass[10pt,5p,times,twocolumn]{elsarticle}

\usepackage{graphicx}
\usepackage{balance}
\usepackage{array,threeparttable}
\usepackage{url}
\usepackage[figuresright]{rotating}
\usepackage{caption}
\usepackage{amsmath}
\usepackage{amssymb}
\usepackage{booktabs}
\usepackage{multirow}
\usepackage{algorithm}
\usepackage{algpseudocode}
\usepackage{enumitem}
\usepackage{placeins}
\usepackage{float}
\usepackage{hyperref}
\usepackage{xcolor}
\usepackage{titlesec}

\hypersetup{hidelinks}

\titleformat*{\section}{\normalsize\bfseries}
\titleformat*{\subsection}{\normalsize\bfseries\itshape}
\titleformat*{\subsubsection}{\normalsize\bfseries\itshape}

\setlist[itemize]{leftmargin=2em,itemsep=0.2em,topsep=0.2em}
\setlist[enumerate]{leftmargin=2em,itemsep=0.2em,topsep=0.2em}

\makeatletter
\def\ps@pprintTitle{%
  \let\@oddhead\@empty
  \let\@evenhead\@empty
  \let\@oddfoot\@empty
  \let\@evenfoot\@empty}
\makeatother

\journal{Journal of Network and Computer Applications}
\biboptions{sort&compress}

\begin{document}
\begin{frontmatter}

\title{\textbf{TraceGrant: A Contract-Governed Security Framework for the Task-Effect Lifecycle of Networked LLM Agents}}

\author[inst1]{Bohao Liao}
\ead{liaobohao@stu.xidian.edu.cn}

\author[inst2]{Jingchao Wang}
\ead{wangjc.2000@tsinghua.org.cn}

\author[inst3]{Qipeng Song}
\ead{qpsong@xidian.edu.cn}

\author[inst3]{Jin Cao}
\ead{jcao@xidian.edu.cn}

\author[inst1]{Jieling Wang}
\ead{jlwang@xidian.edu.cn}

\author[inst2]{Boyu Deng\corref{cor1}}
\ead{kdydby2014@sina.com}

\address[inst1]{School of Telecommunication Engineering, Xidian University, Xi'an, Shaanxi 710071, China}
\address[inst2]{National Key Laboratory of Multi-domain Data Collaborative Processing and Control, Beijing 100141, China}
\address[inst3]{School of Cyber Engineering, Xidian University, Xi'an, Shaanxi 710071, China}

\cortext[cor1]{\raggedright Corresponding author.}

\begin{abstract}

Networked large language model (LLM) agents retrieve information from email, cloud storage, calendars, transaction platforms, and Web services to complete multistep tasks that produce persistent external effects. The same content needed for legitimate execution may also contain indirect prompt injections that redirect tool use, alter sensitive arguments, or disrupt task completion. Existing defenses mainly constrain untrusted content or individual tool calls, leaving user intent, runtime evidence, realized effects, and task completion insufficiently connected. We present TraceGrant, a security framework that governs the task-effect lifecycle of networked LLM agents through an explicit Contract. Before execution, TraceGrant establishes a task-effect boundary from the trusted user request. During execution, admitted evidence can instantiate only authority already established by the Contract. After execution, task completion is verified against actual tool results. Across 949 AgentDojo and 400 Agent Security Bench attack cases under fixed benchmark settings, TraceGrant recorded no attack successes while retaining utility under attack rates of 77.32\% and 83.00\%, respectively. We further evaluate TraceGrant through white-box defense-aware attacks, Contract quality analysis, stage ablations, targeted stress tests, and runtime overhead measurements. The results show that TraceGrant provides a unified governance layer that connects trusted user intent, runtime evidence, concrete tool execution, and verified task completion.

\end{abstract}


\end{frontmatter}

\section{Introduction}

Large language model (LLM) agents are increasingly able to act rather than merely respond. Through tool interfaces, an agent can retrieve information from email, messaging platforms, cloud storage, calendars, databases, payment systems, and Web APIs, coordinate multiple steps, and produce persistent effects in external systems. ReAct, Toolformer, ToolLLM, and related agent frameworks have established this pattern of reasoning, tool use, and observation-driven execution\cite{yao2023react,schick2023toolformer,qin2024toolllm,dibia2024autogenstudio}. As agents gain access to consequential services, their security depends increasingly on whether the effects they produce remain aligned with the authority granted by the user.

This dependence becomes difficult in networked environments because legitimate execution often requires untrusted runtime content. An agent may need to read an email, shared document, Web page, search result, or tool output before it can determine the arguments of a later action. The same content can contain an indirect prompt injection that redirects tool use, substitutes a recipient or account, expands a resource scope, or triggers an additional operation\cite{greshake2023indirect,liu2024promptinjection,yi2025bipia,zhan2024injecagent}. Since the agent already holds legitimate credentials and tool access, such content can influence real external effects rather than merely change generated text.

Consider a user who asks an agent to identify the project owner from a shared workspace and send that person a report. The sending operation is authorized by the user, but the recipient is unknown when the request is issued and must be discovered at runtime. A workspace query may return the correct recipient together with injected text that requests another address or an additional action. The returned data is necessary for task completion, yet its presence in the agent context should not determine the scope of the user's authority. The central security question is therefore whether runtime information can supply values needed by an authorized task without acquiring authority to redefine the effects of that task.

Existing defenses address important parts of this problem through information flow control and isolation, policy based tool authorization, and semantic task alignment\cite{debenedetti2025camel,costa2025fides,wu2024isolategpt,shi2025progent,wang2026agentspec,jia2025taskshield}. These approaches constrain how untrusted content propagates or determine whether a proposed action satisfies a policy or user goal. A networked task, however, evolves across multiple observations and effects. Its execution also depends on which evidence supports a dynamic argument, whether the corresponding task step is currently authorized, how much execution authority remains, and whether the native tool actually produced the required result. Securing an individual message or tool call therefore leaves a broader task state that must remain consistent from the original request through runtime execution and completion.

This setting creates three requirements for task-effect security. First, the user request must be translated into an explicit effect boundary that captures permitted operations, required steps, critical arguments, and execution limits while preserving legitimate runtime flexibility. Second, runtime evidence must be able to instantiate unresolved arguments within that boundary while preserving their provenance, task context, and permitted scope. Third, authorization must remain distinct from actual execution and task completion, so that tool failure, argument drift, unmet postconditions, or partial execution cannot be accepted as successful completion.

We address these requirements with TraceGrant, a Contract governed security framework for the task-effect lifecycle of networked LLM agents. TraceGrant treats runtime data as evidence for existing authority rather than as a source of new authority. Before execution, a Semantic Contract Compiler derives a Permission, Obligation, Evidence, and Completion (POEC) Contract from the trusted user request and admitted tool schemas. The Contract defines permitted effects, required obligations, evidence rules for dynamic arguments, and observable completion conditions. During execution, an Obligation Ledger tracks task state, while admitted evidence, typed argument proofs, call budgets, and a deterministic policy decision and enforcement path govern each proposed effect. After execution, an Effect Receipt binds authorization to the actual native result, and a Final Answer Gate accepts task completion only after the required obligations have been verified.

We evaluate TraceGrant on AgentDojo and Agent Security Bench (ASB). Under the fixed benchmark settings, no attack succeeded across 949 AgentDojo attack cases or 400 ASB attack cases, while utility under attack reached 77.32\% and 83.00\%, respectively. Across all four foundation models used for task execution, no attack success was observed under the fixed benchmark settings. We further evaluate the framework through white-box defense-aware attacks, Contract quality analysis, stage ablations, targeted runtime and completion stress tests, and runtime overhead measurements. Together, these experiments examine the overall security of TraceGrant and the distinct roles of Contract establishment, runtime authorization, and task closure backed by execution results.

This paper makes three contributions:
\begin{enumerate}
  \item \textbf{Task-effect lifecycle modeling and Contract establishment.} We model agent security across authority establishment, runtime argument instantiation, external effects, and task completion. The POEC Contract represents permitted effects, required obligations, evidence rules, and completion conditions, while deterministic static analysis validates the Contract before it becomes executable authority.
  \item \textbf{Runtime authorization bounded by evidence.} TraceGrant combines evidence admission, active obligation matching, typed argument proofs, and call budgets to bind dynamic arguments to qualified runtime evidence. Each Effect Certificate is single use and connects the Contract, current obligation, canonical arguments, and supporting evidence to a concrete external action.
  \item \textbf{Execution verification and task closure.} TraceGrant verifies the actual tool, arguments, execution result, and observable postconditions after each authorized effect. Effect Receipts update task state from native execution, while the Final Answer Gate links verified effects to completion of the overall task.
\end{enumerate}

\section{Related Work}

Existing defenses for LLM agents that interact with external tools mainly address three aspects of agent security: limiting the influence of untrusted content, controlling access to external tools, and aligning agent actions with the user's task. TraceGrant builds on these directions but focuses on the task state that connects user authority, runtime evidence, concrete tool execution, and verified completion. We review the most relevant work from these perspectives.

\subsection{Indirect Prompt Injection and Context Isolation}

Indirect prompt injection arises when trusted user instructions and untrusted external content enter the same execution context. Such attacks have been demonstrated through Web pages, email, shared files, tool outputs, and other data sources used by LLM agents\cite{greshake2023indirect,liu2024promptinjection,yi2025bipia,zhan2024injecagent}. Because these sources often contain information required for legitimate task execution, defenses must control how external content influences later reasoning and actions.

CaMeL separates trusted control flow from untrusted data flow and uses capabilities to restrict how untrusted values propagate into later operations\cite{debenedetti2025camel}. FIDES tracks confidentiality and integrity labels through the agent planner and tool result channels, enforcing information flow policies before consequential actions\cite{costa2025fides}. IsolateGPT uses execution isolation to separate applications and sessions and mediates communication across capability domains\cite{wu2024isolategpt}. These systems reduce the influence of untrusted information before or during tool use.

Task Shield addresses the problem from the perspective of task alignment. It evaluates whether instructions and proposed tool calls contribute to the goal specified by the user and rejects actions that lack sufficient task relevance\cite{jia2025taskshield}. TraceGrant complements these approaches by tracking when runtime information becomes admissible evidence for a concrete task effect. Its authorization decision is tied to the current task obligation, the argument supported by that evidence, and the authority available at the corresponding execution state.

\subsection{Tool Authorization and Behavioral Specifications}

Another line of work places an enforcement layer between the agent and external tools. Progent uses a dedicated policy language to express fine-grained permissions and evaluates proposed tool calls deterministically\cite{shi2025progent}. AgentSpec uses customizable behavioral specifications to describe actions that are allowed, prohibited, or permitted under particular conditions\cite{wang2026agentspec}. These systems follow the broader tradition of runtime policy enforcement, where security decisions are made outside the model before consequential actions are executed\cite{schneider2000enforceable}.

TraceGrant extends this form of authorization with persistent task state. Each proposed effect is associated with an active obligation, evidence supporting its authority-bearing arguments, and the remaining execution budget. The Contract and Obligation Ledger therefore connect a local authorization decision to the evidence and task progress that justify it. This allows execution authority to be established, instantiated, and consumed consistently throughout a multistep task.

\subsection{From Runtime Authorization to Task Completion}

Runtime authorization and runtime verification address different stages of system execution. Runtime policy enforcement determines whether a proposed action satisfies a security specification, while runtime verification checks concrete execution against expected behavior as the system runs\cite{schneider2000enforceable,leucker2009runtime}. For LLM agents that interact with external services, these stages are followed by another decision: whether the observed tool results are sufficient to declare the overall task complete.

TraceGrant connects these stages through a shared task state. Authorization establishes that a proposed effect conforms to the current Contract. Native execution records what actually occurred, and an Effect Receipt verifies the realized tool, arguments, execution result, and observable postconditions. Task completion is reached only after the required obligations have been closed. In this way, TraceGrant connects call authorization, execution verification, and task completion within a continuous governance process.

\section{Problem Formulation}

We first define the system and task effects of a networked LLM agent, including the relationship among the user request, runtime network evidence, and external tool execution. We then specify the adversary and derive four security objectives: task-effect confinement, evidence-constrained instantiation, task-state and ordering consistency, and execution-backed completion.

\subsection{System Model and Task Effects}

\begin{figure*}[!t]
  \centering
  \includegraphics[width=0.99\textwidth,trim=167bp 210bp 68bp 73bp,clip]{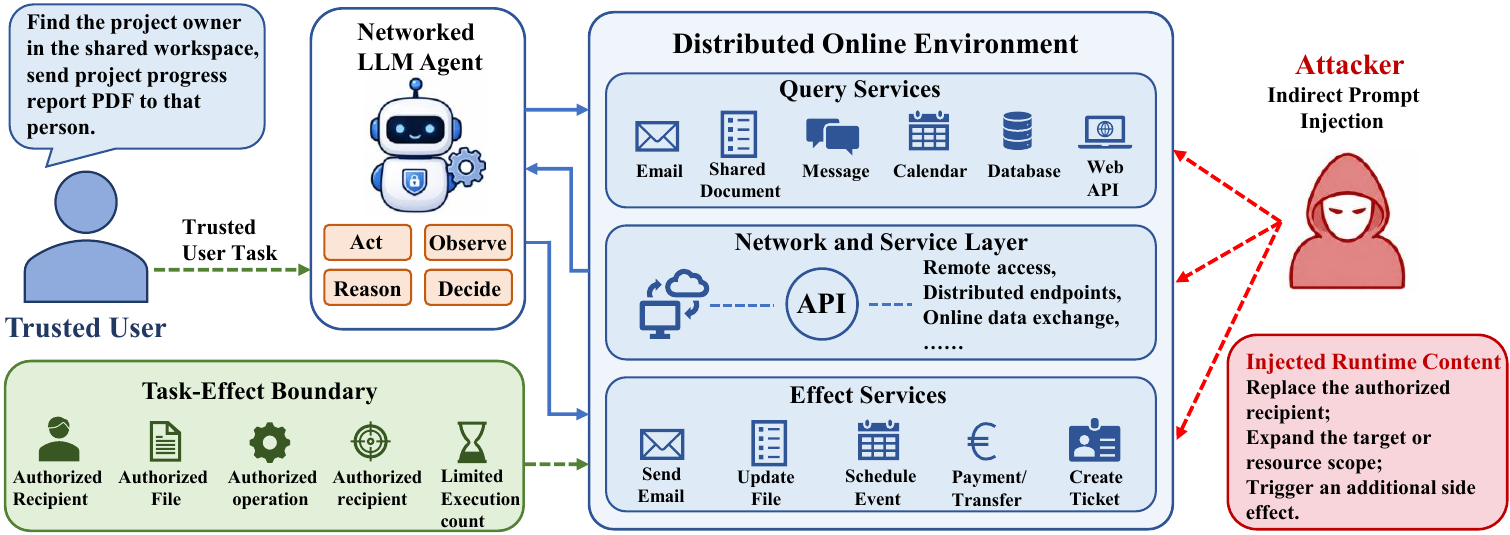}
  \caption{A networked LLM agent crosses service boundaries where untrusted content can influence task effects.}
  \label{fig:networked-agent-scenario}
\end{figure*}

We consider an LLM agent that uses tools, connects to multiple network services, and acts on behalf of a user to complete multistep tasks. The system consists of a trusted user, an LLM agent, an agent runner, tool services accessible over the network, and external content providers. The user submits a natural language request. Based on that request, the execution history, and runtime observations, the agent selects tools that the runner invokes against remote services such as email, messaging, cloud storage, calendars, databases, payment platforms, and Web APIs. Fig.~\ref{fig:networked-agent-scenario} illustrates the task-effect boundary and the threat of indirect prompt injection during retrieval and execution across services.

\noindent\emph{Task effects.} We divide tools into query tools and effect tools. Query tools read external state and return runtime observations for subsequent reasoning. Effect tools send information, modify resources, create calendar entries, initiate transactions, or otherwise change external state. A \emph{task effect} is an observable, persistent, or operationally consequential change outside the agent caused by a tool call. At each discrete execution step, the agent proposes a target tool and a set of arguments. We distinguish \emph{authority-bearing arguments}, which determine the effect's object, resource, quantity, or scope, from \emph{payload arguments}, which carry business content such as a message body, note, or description without directly determining the authorization boundary. The two roles are treated differently during security evaluation.

Typical authority-bearing arguments include an email recipient, destination account, transfer amount, file path, calendar attendee, or resource scope. An attacker can change the realized effect by controlling these arguments even if the tool type remains unchanged. Sending a specified report may be authorized, for example, but substituting a recipient controlled by the attacker still produces an unauthorized outcome. A security mediator must therefore determine not only whether a tool may be called, but also which object and scope the effect may reach and which evidence supports each authority-bearing argument.

For a trusted user request $u$, let $\mathcal{A}(u)$ denote its authorized effect boundary. This boundary covers effect types, target classes, resource scopes, constraints on critical arguments, and maximum invocation counts. Some constraints follow directly from the request; others can be instantiated only through runtime queries. In the running example, the user authorizes sending a specified report, while a constrained workspace query supplies the recipient. The example combines cross-service retrieval, a dynamic runtime recipient, the possibility of poisoned content, and a persistent sending effect. The same abstraction applies to network operations and cloud-resource control, although these are not the primary experimental domains in this paper.

Our objective is to preserve the informational role of runtime data while separating it from effect authority: external data may supply values needed to complete the task, whereas effect authority remains rooted in the trusted user request.

\subsection{Threat Model}

\noindent\emph{Adversary capabilities.} We consider an attacker who can control or influence external content read by the agent. The attacker may embed malicious instructions in email, Web pages, shared files, chat messages, cloud documents, or tool outputs. When the agent reads this material during a legitimate task, the injected instruction enters the model context alongside ordinary business data and may induce candidate actions that depart from the user's request.

The attacker may attempt to influence tool selection, authority-bearing arguments, execution order, invocation count, or the final task claim. We focus on substitutions of recipients, accounts, amounts, or resources from unauthorized sources; expansion of target sets or resource scope; effect execution before required prerequisites have completed; repeated consumption of single-use authority; and success claims made after execution failure, argument mismatch, or partial completion. We also model benign runtime faults, including native tool failure, disagreement between authorized and actual arguments, and unmet observable completion conditions.

The attacker cannot modify the trusted user request or the frozen tool registry used for Contract compilation, and cannot bypass the protected tool boundary. Runtime tool additions or schema changes are treated as untrusted unless admitted against this registry.

The adversary may poison fields within an external object that is otherwise admissible as task evidence. TraceGrant does not equate evidence admissibility with semantic authenticity; its conditional guarantee assumes authentic authority-bearing evidence fields, while Section~\ref{sec:defense-aware-attacks} evaluates attacks that violate this condition.

\subsection{Security Objectives}

Given this system and threat model, a defense should preserve legitimate dynamic execution while satisfying the following task-effect security objectives.

TraceGrant focuses on the authorization and completion integrity of external effects. Confidentiality constraints over payload content can be enforced by complementary information flow or isolation mechanisms alongside the task-effect controls considered here.

\noindent\emph{Task-effect confinement.} Every invocation of an effect tool must pass through one protected execution boundary. The realized action must remain within the effect type, target scope, argument constraints, and call budget authorized by the user request. An effect introduced only by external content must never become executable.

\noindent\emph{Evidence-constrained instantiation.} Runtime data may fill only a dynamic argument slot already present in the task authorized by the user, and the value must satisfy the expected type, permitted source, query context, and constraints at the point of use. Only admitted evidence may instantiate an unresolved part of the boundary established before execution, and the resulting runtime authority must preserve or narrow that boundary. Runtime data may not introduce a tool or effect type, broaden a target or resource scope, increase a call budget, or change the required task structure.

\noindent\emph{Task-state and ordering consistency.} An effect in a multistep task must correspond to an outstanding obligation whose prerequisites have been met. Arguments from future steps, effects whose dependencies remain unsatisfied, and repeated calls beyond the remaining budget must not be executable. Within the established boundary, the agent should still be able to issue additional queries, correct arguments, and perform legitimate retries.

\noindent\emph{Execution-backed completion.} Authorization of a candidate action, successful return from a native tool, and completion of the overall task are distinct states. Authorization establishes only that a candidate fits the current effect boundary; native success establishes only that one call returned successfully. A success claim at the task level is acceptable only when the actual tool and arguments match the authorization, observable completion conditions hold, and every hard obligation has verified execution support. This objective prevents tool failure, argument drift, unmet postconditions, partial completion, or a premature claim from being promoted to successful task completion.
\section{TraceGrant Framework}

TraceGrant governs task-effect authority through Contract establishment before execution, runtime authorization, and verification after execution. Runtime data may instantiate authority within an established Contract but cannot create, expand, or replace it.

\subsection{Framework Overview and Design Principles}

\begin{figure*}[!t]
  \centering
  \includegraphics[width=0.98\textwidth,trim=0 75bp 0 0,clip]{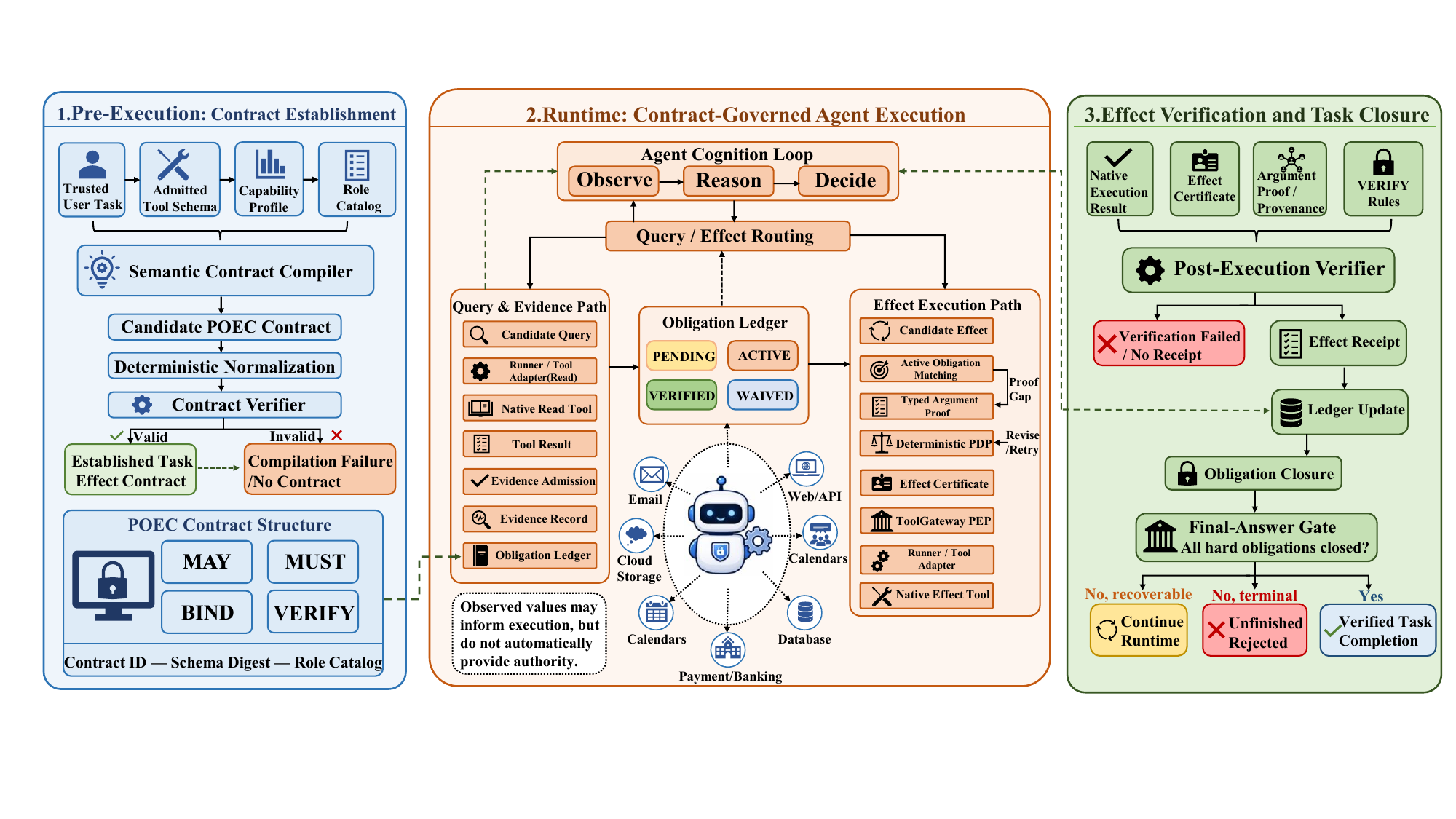}
  \caption{TraceGrant establishes, authorizes, and verifies task effects through a one-way authorization chain.}
  \label{fig:tracegrant-architecture}
\end{figure*}

As shown in Fig.~\ref{fig:tracegrant-architecture}, TraceGrant comprises Contract establishment before execution, runtime authorization bounded by evidence, and verification and task closure after execution. The first stage turns a natural language request into a checkable effect boundary. The second permits qualified evidence to instantiate reserved dynamic arguments and evaluates each candidate effect individually. The third verifies the actual tool call and its result; only a verified effect may advance task state. An ordinary network observation does not itself confer effect authority, and an authorized candidate does not itself establish task completion.

In the running example, the user authorizes the effect of sending a specified report. The workspace result instantiates the recipient within the Contract boundary, and verification of the native send result closes the corresponding obligation. A network result can therefore supply a dynamic argument required by an authorized effect, but cannot change the recipient, file, or operation scope on its own.

\subsection{Task-Effect Contract Establishment Before Execution}

\subsubsection{Semantic Contract Compilation}

Before the agent reads any runtime network content, the pre-execution stage derives a task-effect Contract from the trusted user request and a frozen tool registry. The semantic compiler receives the request $u$, the admitted tool-schema set $\Sigma$, capability profiles $\mathcal P$, and the schema-derived argument role catalog $R_{\Sigma}$, and produces a typed POEC candidate Contract:
\begin{equation}
\widetilde C=F_{\mathrm{sem}}(u,\Sigma,\mathcal P,R_{\Sigma})
=\langle \mathrm{MAY},\mathrm{MUST},\mathrm{BIND},\mathrm{VERIFY}\rangle .
\label{eq:semantic-contract-compilation}
\end{equation}

Only schemas admitted against the frozen registry enter $\Sigma$; runtime additions or modified schemas remain outside the Contract compilation context.

A candidate becomes an established Contract only after deterministic normalization and verification; invalid candidates produce no executable Contract.
Semantic compilation balances minimal authority with legitimate runtime flexibility. Tools, objects, resources, and arguments explicitly named by the user become direct effect constraints. A recipient, account, time, or resource that must be discovered later remains a dynamic slot associated in advance with an evidence obligation, permitted source, and derivation rule. Required queries, effects, and outputs form a dependency-aware obligation structure. Call budgets follow the user's request and the minimum work required by the task, while completion conditions are expressed over native tool status, actual arguments, and observable results. The candidate Contract therefore limits the effects the task may produce without precluding the network evidence needed to complete it.

The four components jointly capture the task-effect boundary and give the Permission--Obligation--Evidence--Completion (POEC) Contract its name. MAY corresponds to Permission and specifies permitted tools and operations, target scope, argument constraints, and call budgets. MUST corresponds to Obligation and organizes evidence acquisition, external effects, and output requirements into an obligation graph. BIND corresponds to Evidence and defines the permitted source, expected type, evidence constraints for a field, and permitted derivation of each authority-bearing argument. For evidence sets, BIND additionally specifies a deterministic selection rule and the structured fields on which the rule may depend; an element selected solely by the LLM cannot constitute an authorization basis. VERIFY corresponds to Completion and states the requirements for the tool, arguments, execution status, and postconditions that must hold after an effect. Together, these elements connect authorization to runtime progress and completion: MAY sets the upper bound on authority; MUST places an action in the task structure; BIND specifies how a dynamic argument obtains a valid authorization basis from runtime evidence, including both its provenance and the admissibility of the referenced record or field; and VERIFY determines when concrete execution counts as task progress.

The argument role catalog separates authority-bearing arguments, which determine the effect's object, resource, or scope, from payload arguments that primarily carry business content. Recipients, accounts, files, amounts, and resource scopes require explicit BIND rules. Payload arguments, including message bodies, descriptions, and notes, are tracked separately from authority-bearing arguments. Their provenance is recorded to distinguish content explicitly authorized by the trusted user request from content influenced by untrusted runtime observations. Tainted payloads without explicit user authorization are subject to runtime risk screening before effect authorization. These role semantics refine permission for a tool into a task-effect boundary at the argument level.

\subsubsection{Static Contract Analysis}

The candidate Contract is treated as a typed intermediate representation rather than executable authority. After deterministic normalization, the Contract Verifier performs static analysis over this representation before any runtime effect is permitted. Let $C'=\operatorname{Normalize}(\widetilde C)$. Its acceptance condition is
\begin{equation}
\small
\begin{aligned}
\operatorname{StaticAccept}(C')={}&
\operatorname{WellFormed}(C')\\
&\land\operatorname{ControlSafe}(C')\\
&\land\operatorname{FlowSafe}(C')\\
&\land\operatorname{ScopeSafe}(C')\\
&\land\operatorname{VerifySafe}(C').
\end{aligned}
\label{eq:static-contract-acceptance}
\end{equation}
\emph{WellFormed} checks consistency of tool schemas, argument types, and references. \emph{ControlSafe} checks obligation dependencies, reachability, and illegal cycles. \emph{FlowSafe} ensures that every authority-bearing argument has an admissible BIND source and derivation. \emph{ScopeSafe} checks the internal consistency of effect scopes, argument constraints, and call budgets. \emph{VerifySafe} ensures that reachable effects have observable verification conditions supported by the schema.

Algorithm~\ref{alg:static-contract-analysis} summarizes this deterministic analysis against $\Sigma$, the role catalog, and the observability catalog.

\begin{algorithm}[t]
\caption{Deterministic static analysis validates a candidate POEC Contract before it becomes executable authority.}
\label{alg:static-contract-analysis}
\begin{algorithmic}[1]
\Require $\widetilde C,\Sigma,R_{\Sigma},\mathcal O_{\Sigma}$
\Ensure \textsc{Accept} or \textsc{Reject}
\Procedure{StaticAnalyze}{}
  \State $C'\gets\Call{Normalize}{\widetilde C}$
  \If{$\neg\Call{SchemaRefTypeCheck}{C',\Sigma}$}
    \State \Return \textsc{Reject} \Comment{WellFormed}
  \EndIf
  \State $(G_h,R)\gets\Call{DependencyReachability}{C'.\mathrm{MUST}}$
  \If{$\neg\Call{Acyclic}{G_h}\lor\Call{HardNodes}{G_h}\nsubseteq R$}
    \State \Return \textsc{Reject} \Comment{ControlSafe}
  \EndIf
  \State $E\gets\Call{ReachableEffects}{C',R}$
  \If{$\neg\Call{CheckBindings}{E,C',R_{\Sigma},\Sigma}$}
    \State \Return \textsc{Reject} \Comment{FlowSafe}
  \EndIf
  \If{$\neg\Call{CheckScope}{C',R_{\Sigma},\Sigma}\lor
      \neg\Call{CheckVerification}{E,C',\mathcal O_{\Sigma},\Sigma}$}
    \State \Return \textsc{Reject} \Comment{ScopeSafe or VerifySafe}
  \EndIf
  \State \Return \textsc{Accept}
\EndProcedure
\end{algorithmic}
\end{algorithm}

If accepted, the normalized representation $C'$ becomes the established Contract $C$ and enters the Obligation Ledger; otherwise, no executable Contract is created.

An established Contract feeds two consistent paths. The full POEC Contract and currently eligible obligations enter the agent's execution context alongside the trusted request, tool schemas, and history, allowing the agent to plan queries, construct arguments, and propose actions. The same Contract is written to the Obligation Ledger and enforced by the policy decision point (PDP), policy enforcement point (PEP), and post-execution verifier. The agent plans against the same machine-checkable specification that the mediator enforces. The Contract remains stable during execution; observations and active obligations evolve, while the Ledger records admitted evidence, consumed budget, and completed steps.
\subsection{Runtime Evidence-Bounded Effect Authorization}
\label{sec:runtime-effect-authorization}

Static analysis validates the structure and internal consistency of the candidate Contract before execution, whereas runtime enforcement evaluates concrete evidence, argument values, obligation states, and remaining budgets.

Once the Contract is established, the agent resumes its ordinary observe--reason--act loop. TraceGrant routes candidate requests through either a query path or an effect path, while the Obligation Ledger maintains dynamic task state. Each obligation is PENDING, ACTIVE, VERIFIED, or WAIVED. A PENDING obligation becomes ACTIVE when its dependencies are satisfied. Admitted evidence or a valid execution result can move the corresponding obligation to VERIFIED. Completion of one obligation makes its successors eligible according to the dependency relation in MUST. At each step, the agent receives active obligations and tool results and may propose another query, an effect candidate, or a final answer.

The dynamic Ledger state at step $t$ is
\begin{equation}
\begin{aligned}
L_t&=\langle E_t,S_t,B_t,K_t\rangle,\\
S_t&:\mathcal{O}_C\rightarrow\mathcal{S}_O,\\
\mathcal{S}_O&=\{\mathrm{PENDING},\mathrm{ACTIVE},\\
&\qquad\mathrm{VERIFIED},\mathrm{WAIVED}\}.
\end{aligned}
\label{eq:ledger-state}
\end{equation}
Here, $E_t$ is the set of admitted Evidence Records, $\mathcal{O}_C$ is the Contract obligation set, $\mathcal{S}_O$ is the obligation-state alphabet, $S_t$ is the obligation-state mapping, $B_t$ is the remaining-budget vector, and $K_t$ is the Effect Certificate lifecycle registry recording issued and consumed certificates. Accordingly, $\operatorname{Reach}(C)$ in Section~4.5 denotes the Ledger states reachable from the initial state $L_0$ under the Contract-governed transitions.

A query result first enters the agent context as an ordinary runtime observation. Evidence Admission promotes it to task evidence only if it satisfies an active evidence obligation and matches the required source tool, query conditions, field type, and evidence constraints defined by BIND. Let $\operatorname{state}_t(o_i)$ denote the state of obligation $o_i$ at time $t$:
\begin{equation}
\begin{aligned}
\operatorname{Admit}_{C}(r_t,o_i,L_t)=1 \Longleftrightarrow{}&
\operatorname{state}_t(o_i)=\mathrm{ACTIVE}\\
&\land \operatorname{SourceMatch}_{C}(r_t,o_i)\\
&\land \operatorname{QueryMatch}_{C}(r_t,o_i)\\
&\land \operatorname{TypeMatch}_{C}(r_t,o_i)\\
&\land \operatorname{IntegrityMatch}_{C}(r_t,o_i).
\end{aligned}
\label{eq:evidence-admission}
\end{equation}
IntegrityMatch checks whether the returned record or field satisfies the evidence constraints specified by BIND using the structured attributes available from the corresponding tool result. A result that fails this check remains available as an ordinary runtime observation but cannot be recorded as evidence for an authority-bearing argument. The admitted result is written to the Ledger together with references among the source tool, query arguments, returned field, and evidence obligation. Other observations remain available for environmental understanding and reasoning, but only admitted evidence may support an authority-bearing recipient, account, amount, file, or resource scope. Once the evidence obligation closes, its successor effect obligations in MUST can become ACTIVE.

When the agent proposes an effect $a_t^{e}=(\tau_t,\theta_t)$, TraceGrant first matches it to an active effect obligation and constructs a typed proof $\Pi_t$ for every authority-bearing argument. A typed proof may cite a constant in the trusted request or an admitted Evidence Record in the Ledger. For every authority-bearing argument, the proof establishes both the provenance of the value and the admissibility of the supporting record or field. Field extraction, type conversion, and numerical computation must follow a derivation allowed by BIND. Set selection is valid only when a trusted deterministic selector applies the BIND-defined rule and produces a unique result. Zero or multiple matches cannot support a valid argument proof. For a selected value, the typed argument proof additionally binds the candidate set, the BIND-defined selector, and the selected record. The PDP accepts the value only when the selection can be deterministically reproduced from the admitted evidence. Before validating an existing one-shot grant, the PDP performs provenance-aware payload screening. Let $H$ indicate that a payload originates from untrusted external content, is not explicitly authorized by the trusted user request, and matches a high-confidence risk pattern, including active scripts, command-execution content, instruction-hijacking markers, or explicit credential material. The PDP then makes a stateful decision over the payload-screening result, permitted scope, active obligation, argument proofs, and remaining budget. For compactness, $M$, $O$, $P$, and $B$ denote the MAY, active-obligation, PROOF, and BUDGET conditions:
\begin{equation}
\small
\operatorname{Decision}_{C}\!\left(a_t^{e},L_t\right)=
\left\{
\begin{array}{@{}l@{}l@{}}
\mathrm{STEP\_UP\_AUTH}, & H,\\[1mm]
\mathrm{ALLOW}, & \neg H\land M\land O\land P\land B,\\[1mm]
\mathrm{PROOF\text{-}GAP}, & \neg H\land M\land O\land B\land\neg P,\\[1mm]
\mathrm{DENY}, & \text{otherwise}.
\end{array}
\right.
\label{eq:effect-decision}
\end{equation}
STEP\_UP\_AUTH places a candidate effect in a pending authorization state, during which the PEP withholds native tool execution. A trusted authentication channel presents the exact tool and canonical arguments to the user. After re-authentication and approval, the system issues a short-lived, one-shot grant bound to the Contract, session, tool, and argument digest; the PDP/PEP then re-evaluates the candidate under that grant. Any change to the arguments initiates a new authorization decision, and a sanitized alternative enters the complete pipeline as a new candidate. ALLOW means that the candidate lies within MAY, advances an active obligation, has a valid proof for every authority-bearing argument, and remains within budget. PROOF-GAP means that the effect type and task step fit the Contract, but at least one dynamic argument lacks the evidence required by BIND. TraceGrant returns the missing slot, source, or derivation to the agent so that it can query for evidence or revise the candidate. DENY covers candidates outside MAY, candidates unrelated to an active obligation, scope-expanding arguments, and exhausted budgets. A denied request never reaches the native effect tool.

After an ALLOW decision, the PDP issues a one-shot Effect Certificate:
\begin{equation}
\kappa_t=\left\langle
id_C,o_i,\tau_t,\operatorname{Canon}(\theta_t),\Pi_t,b_t
\right\rangle .
\label{eq:effect-certificate}
\end{equation}
Here, $id_C$ identifies the Contract, $o_i$ is the matched active obligation, $\operatorname{Canon}(\theta_t)$ contains canonical tool arguments, $\Pi_t$ contains the argument proofs, and $b_t$ records the budget state for this call. The Effect Certificate narrows abstract task authority to one exact execution and keeps the authorized object, argument values, evidence provenance and integrity qualification, and task step bound together.

In our single-trust-domain implementation, an Effect Certificate remains a logical object inside the protected mediator, whose atomic issuance, validation, and consumption provide one-shot semantics. With network-separated PDP and PEP components, it must instead be an authenticated, time-bounded, replay-resistant capability that is consumed atomically with the Ledger update; the guarantees do not depend on a particular cryptographic construction.

The ToolGateway PEP enforces the PDP's decision. It dispatches a call to the protected runner only when the requested tool, canonical arguments, and current obligation exactly match a valid Effect Certificate:
\begin{equation}
\small
\begin{aligned}
\operatorname{Dispatch}(a_t^{e})\Longleftrightarrow{}&
\operatorname{ValidCert}(\kappa_t,L_t)\\
&\land\operatorname{Match}(a_t^{e},\kappa_t).
\end{aligned}
\label{eq:pep-dispatch}
\end{equation}
The Effect Certificate is consumed when the corresponding call is dispatched and cannot authorize a second effect. A repeated operation or legitimate retry must return to the PDP under the updated obligation state and remaining budget. DENY and PROOF-GAP produce no Effect Certificate, so the PEP does not invoke the native effect tool.

Together, Equations~\eqref{eq:effect-decision}--\eqref{eq:pep-dispatch} ensure that an effect is dispatched only after an \texttt{ALLOW} decision and only when the requested tool and canonical arguments exactly match a valid one-shot Effect Certificate. Runtime execution is therefore confined to the established Contract under complete PEP mediation.
This sequence forms a continuous Contract-governance loop. Admitted query results add evidence, the Ledger advances obligations, the agent proposes new candidates from the updated context, and the PDP and PEP authorize each external effect separately. When evidence is missing, the agent can follow a PROOF-GAP response with another query or a corrected argument. When an effect fails transiently, the agent can request a retry within the original Contract and remaining budget. Runtime data participates in a legitimate task only through evidence admission, argument proof, obligation matching, PDP authorization, and PEP dispatch. The pre-execution Contract and dynamic Ledger jointly constrain the type, object, scope, order, and number of effects throughout execution.

\subsection{Receipt-Backed Effect Verification and Task Closure}

An Effect Certificate establishes that a candidate conforms to the current authorization boundary and task state; dispatch does not establish that the effect occurred. A native call may fail, the arguments received by the tool may differ from those authorized, or a successful return may still leave a task postcondition unsatisfied. TraceGrant therefore distinguishes authorization, successful native execution, and completion of the overall task, and requires evidence for every transition among them.
After the native tool returns, TraceGrant jointly checks the realized call, Effect Certificate, argument provenance, and VERIFY rules. The validity of one effect is
\begin{equation}
\begin{aligned}
\operatorname{EffectValid}_C(y_t,\kappa_t,L_t)={}&
\operatorname{CertMatch}(y_t,\kappa_t)\\
&\land\operatorname{ExecSuccess}(y_t)\\
&\land\operatorname{PostSatisfied}_C(y_t)\\
&\land\operatorname{ProvenanceValid}(\kappa_t,L_t).
\end{aligned}
\label{eq:effect-verification}
\end{equation}
CertMatch checks that the actual tool and canonical arguments match the Effect Certificate. ExecSuccess requires a valid native execution result. PostSatisfied checks the observable completion conditions in the Contract. ProvenanceValid confirms that the argument proofs underlying the authorization remain valid in the current task state. This rule prevents dispatch, an interface return, or a model-generated statement of success from being treated as proof that the effect completed.

If verification succeeds, TraceGrant creates an Effect Receipt linking the authority encoded in the Effect Certificate to the native result. An Effect Receipt establishes not only that a call occurred, but that its tool, arguments, provenance, and completion conditions satisfy the active obligation. Only a valid Effect Receipt moves an effect obligation from ACTIVE to VERIFIED and enables dependent obligations. Execution failure, argument mismatch, or an unmet postcondition leaves the task incomplete.
Before the agent returns a final answer, TraceGrant checks whether every hard obligation has valid completion support:
\begin{equation}
\small
\begin{aligned}
\operatorname{AcceptFinal}_C(L_t)&\Longleftrightarrow
\forall o_i\in \mathcal{O}_C^H,\\
&\operatorname{state}_t(o_i)\in\{\mathrm{VERIFIED},\mathrm{WAIVED}\}.
\end{aligned}
\label{eq:final-answer-gate}
\end{equation}
VERIFIED means that admitted evidence or a valid Effect Receipt has closed the obligation. WAIVED means that a legitimate exemption specified in the Contract applies. The system accepts a success claim only when every hard obligation is in one of these two states.

The Effect Receipt answers whether one effect completed. The Final-Answer Gate asks whether all verified effects together constitute the required task outcome. Success of one tool call closes only its corresponding obligation and cannot substitute for another required step. Task completion is accepted only when every hard obligation has valid support. TraceGrant thereby lifts effect-level execution verification into task-level completion-integrity enforcement.

\subsection{Conditional Security Guarantees}

To separate enforcement rules from security claims, let $\mathsf{AdmEff}(C,L)$ be the set of canonical effects that satisfy MAY, an active obligation, the BIND argument proofs, and the remaining budget in Ledger state $L$. Let $\mathsf{Exec}(\rho)$ be the effects actually produced by native tools along trace $\rho$. A Contract is semantically sound for user request $u$ when every effect admissible in every reachable Contract state remains inside the user-authorized boundary:
\begin{equation}
\begin{aligned}
\operatorname{Sound}(C,u)\Longleftrightarrow{}&
\forall L\in\operatorname{Reach}(C),\\
&\mathsf{AdmEff}(C,L)\subseteq\mathcal{A}(u).
\end{aligned}
\label{eq:contract-soundness}
\end{equation}

Section~\ref{sec:contract-quality-experiment} empirically assesses this semantic-soundness assumption by comparing generated Contracts with independently constructed reference Contracts.

Under semantic Contract soundness, semantic authenticity of the admitted evidence fields used to derive authority-bearing arguments, complete PEP mediation, atomic Certificate and Ledger operations, and exact protected-runner execution, every realized effect remains within the user-authorized boundary $\mathcal{A}(u)$. Network-separated deployments additionally require authenticated, replay-resistant, and time-bounded Effect Certificates.

The same assumptions support completion integrity: a final success claim is accepted only when all non-waived hard obligations are supported by admitted evidence or valid Effect Receipts. Execution failure, argument mismatch, or unsatisfied postconditions therefore cannot support task completion.

\section{Experiments}

We evaluate TraceGrant's security, task utility, and three-stage design. Overall comparisons on AgentDojo and Agent Security Bench are followed by stage ablations and Contract-quality measurements. Targeted stress tests isolate runtime authorization and completion-integrity mechanisms. We further conduct a white-box defense-aware attack experiment in which the attacker knows TraceGrant's defense structure and targets pre-run, runtime, and post-run surfaces. We then quantify runtime overhead before using two end-to-end cases to trace an effect from Contract establishment and evidence binding through execution verification and task closure.

\begin{table*}[!t]
  \centering
  \caption{Overall security and task utility of TraceGrant and the baselines across benchmarks and task-executing models.}
  \label{tab:overall-results}
  \small
  \setlength{\tabcolsep}{3.5pt}
  \renewcommand{\arraystretch}{1.08}
  \begin{tabular*}{\textwidth}{@{\extracolsep{\fill}}>{\raggedright\arraybackslash}p{0.19\textwidth}rrr>{\raggedright\arraybackslash}p{0.19\textwidth}rrr@{}}
    \toprule
    \multicolumn{4}{c}{AgentDojo} & \multicolumn{4}{c}{ASB} \\
    \cmidrule(lr){1-4}\cmidrule(lr){5-8}
    Method / Model & TSR & UUA & ASR & Method / Model & TSR & UUA & ASR \\
    \midrule
    NoDefense & 85.28\% & 73.10\% & 9.64\% & NoDefense & 84.00\% & 42.00\% & 85.00\% \\
    Progent & 73.62\% & 60.60\% & 1.52\% & Progent & 78.25\% & 68.50\% & 4.00\% \\
    CaMeL & 75.30\% & 76.40\% & 1.48\% & CaMeL & 64.00\% & 54.00\% & 19.00\% \\
    AgentSpec & 55.84\% & 53.30\% & 2.54\% & AgentSpec & 44.25\% & 11.00\% & 10.00\% \\
    IsolateGPT & 72.62\% & 58.10\% & 2.03\% & IsolateGPT & 57.00\% & 16.00\% & 12.50\% \\
    FIDES & 56.91\% & 53.31\% & 0.42\% & FIDES & -- & -- & -- \\
    Task Shield & 68.54\% & 61.79\% & 2.24\% & Task Shield & 63.00\% & 57.00\% & 9.25\% \\
    \midrule
    \multicolumn{4}{l}{\textbf{TraceGrant}} &
    \multicolumn{4}{l}{\textbf{TraceGrant}} \\
    DeepSeek-V4-Flash & 80.71\% & 77.32\% & 0.00\% & DeepSeek-V4-Flash & 80.50\% & 83.00\% & 0.00\% \\
    Gemini 3.6 Flash & 75.66\% & 73.20\% & 0.00\% & Gemini 3.6 Flash & 74.13\% & 70.45\% & 0.00\% \\
    Qwen3.7-Plus & 74.47\% & 73.68\% & 0.00\% & Qwen3.7-Plus & 78.33\% & 64.00\% & 0.00\% \\
    GLM-5.2 & 70.14\% & 70.57\% & 0.00\% & GLM-5.2 & 83.47\% & 82.93\% & 0.00\% \\
    \bottomrule
  \end{tabular*}
\end{table*}

\begin{figure*}[!t]
  \centering
  \includegraphics[width=0.99\textwidth]{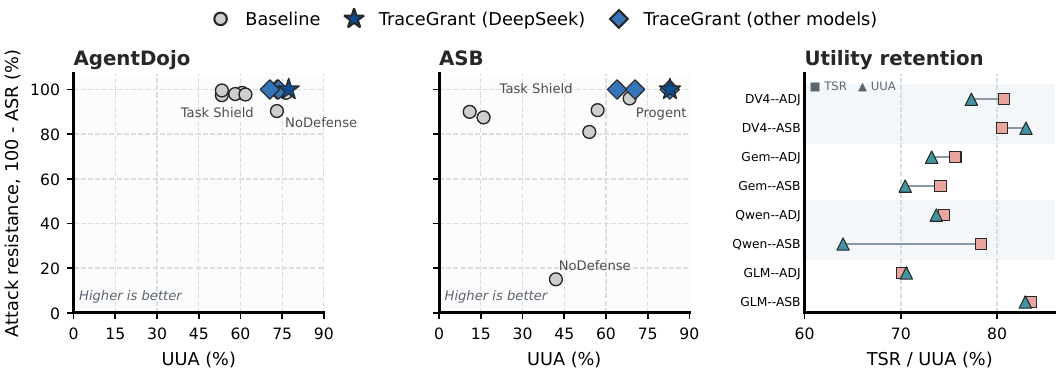}
  \captionsetup{skip=4pt}
  \caption{Security--utility comparison on AgentDojo and ASB, together with cross-model utility retention for TraceGrant.}
  \label{fig:overall-security-utility-profile}
\end{figure*}

\newcommand{\DetailedResultTables}{%
\begin{table*}[!t]
  \centering
  \caption{Stage ablations isolate the contributions of Contract establishment, runtime authorization, and receipt-backed completion.}
  \label{tab:stage-ablation}
  \small
  \setlength{\tabcolsep}{4pt}
  \renewcommand{\arraystretch}{1.08}
  \begin{tabular*}{\textwidth}{@{\extracolsep{\fill}}>{\raggedright\arraybackslash}p{0.31\textwidth}rrrrrr@{}}
    \toprule
    Method & TSR & UUA & ASR & SVR & SOOR & FCR \\
    \midrule
    \textbf{Full TraceGrant} & \textbf{80.71\%} & \textbf{77.32\%} & \textbf{0.00\%} & \textbf{0.00\%} & \textbf{0.00\%} & \textbf{9.78\%} \\
    w/o Pre-execution Specification & 78.95\% & 76.00\% & 0.00\% & 64.19\% & 13.49\% & 42.74\% \\
    w/o Runtime Enforcement & 77.44\% & 75.27\% & 0.76\% & 64.75\% & 2.40\% & 14.81\% \\
    w/o Receipt-backed Completion & 68.42\% & 70.99\% & 0.00\% & 0.00\% & 0.00\% & 16.28\% \\
    Prompt-only Plan & 83.98\% & 74.45\% & 3.27\% & 42.40\% & 10.80\% & 31.70\% \\
    \bottomrule
  \end{tabular*}

\end{table*}

\begin{table}[!t]
  \centering
  \caption{Contract compilation quality is measured against reference task-effect boundaries on AgentDojo.}
  \label{tab:contract-quality}
  \footnotesize
  \setlength{\tabcolsep}{1.0pt}
  \renewcommand{\arraystretch}{1.08}
  \begin{tabular*}{\columnwidth}{@{\extracolsep{\fill}}lrrrrrr@{}}
    \toprule
    Suite & Tasks & VCR & \shortstack{MAY\\F1} & \shortstack{MUST\\F1} & \shortstack{BIND\\Acc.} & \shortstack{VERIFY\\Acc.} \\
    \midrule
    Workspace & 40 & 100.0 & 96.7 & 87.0 & 83.0 & 88.5 \\
    Slack & 21 & 100.0 & 90.0 & 84.1 & 81.2 & 82.6 \\
    Travel & 20 & 100.0 & 100.0 & 98.5 & 100.0 & 100.0 \\
    Banking & 16 & 100.0 & 87.5 & 94.8 & 83.3 & 92.9 \\
    \midrule
    \textbf{Overall} & \textbf{97} & \textbf{100.0} & \textbf{94.4} & \textbf{90.1} & \textbf{84.3} & \textbf{88.4} \\
    \bottomrule
  \end{tabular*}
\end{table}
}

\newcommand{\TargetedResultFigures}{%
\begin{figure*}[!t]
  \centering
  \begin{minipage}[t]{0.485\textwidth}
    \vspace{0pt}
    \centering
    \includegraphics[width=\linewidth]{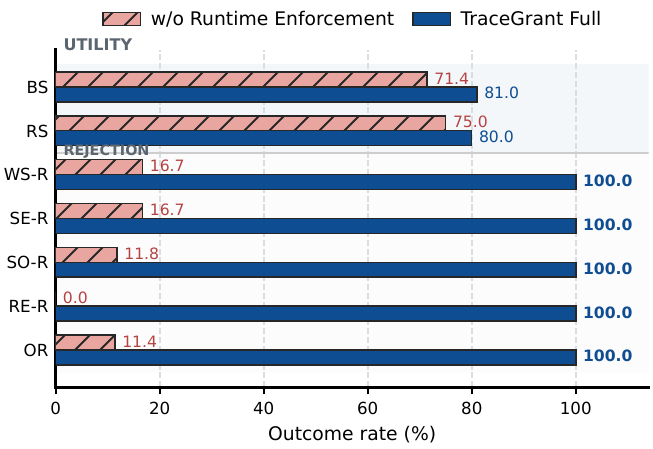}
    \captionof{figure}{Runtime authorization preserves utility while rejecting adversarial candidates at the effect boundary.}
    \label{fig:runtime-enforcement-effects}
  \end{minipage}\hfill
  \begin{minipage}[t]{0.485\textwidth}
    \vspace{0pt}
    \centering
    \includegraphics[width=\linewidth]{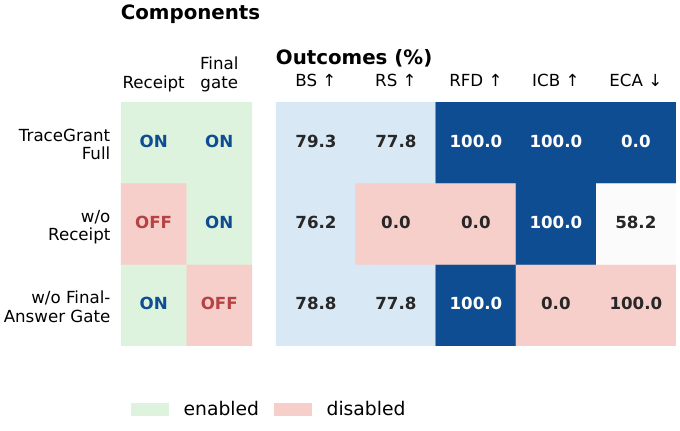}
    \captionof{figure}{Receipt verification and final-answer gating enforce distinct layers of completion integrity.}
    \label{fig:completion-integrity-effects}
  \end{minipage}

\end{figure*}

\begin{figure}[!t]
  \centering
  \includegraphics[width=\columnwidth]{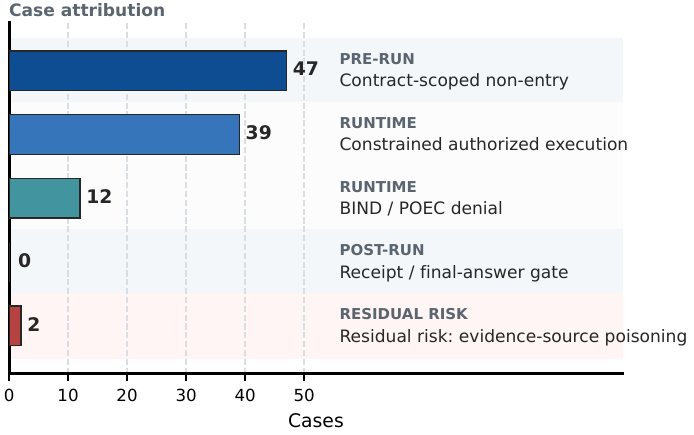}
  \captionsetup{skip=4pt}
  \caption{White-box attack outcomes are attributed to the ordered TraceGrant defense stages on AgentDojo ($n=100$).}
  \label{fig:whitebox-adaptive-layers}
\end{figure}
}

\subsection{Experimental Setup}

\subsubsection{Benchmarks}

We use AgentDojo and Agent Security Bench (ASB)\cite{debenedetti2024agentdojo,zhang2025asb}. Both benchmarks evaluate LLM agents with access to external tools and environment state, but differ in environment organization, task composition, and attack coverage.

\emph{AgentDojo.} AgentDojo contains four interactive environments: Workspace, Slack, Travel, and Banking. They provide 24, 11, 28, and 11 tools; 40, 21, 20, and 16 user tasks; and 14, 5, 7, and 9 injection tasks, respectively. We use the complete task set. The benign setting contains all 97 user tasks. The attack setting pairs every user task with every injection task from the same environment, yielding 949 test cases (hereafter, \emph{cases}): 560 for Workspace, 105 for Slack, 140 for Travel, and 144 for Banking. In each repetition, a method produces 97 benign and 949 attack runs, or 1,046 trajectories. TSR is computed over the 97 benign cases; UUA and ASR are computed over the 949 attack cases. Each environment also supplies an initial state, native tool schemas, a task-utility evaluator, and an attack-goal evaluator.

\emph{ASB.} ASB spans ten domains: system administration, financial analysis, legal services, healthcare, education, psychological counseling, e-commerce, aerospace design, academic search, and autonomous driving. It includes 20 benign tools, 400 attack tools, 51 benign tasks, and 400 attack instructions. We use the complete 400-case benchmark. Each case combines a benign task from the relevant domain with one attack tool and its instruction. In each repetition, every comparison method except FIDES runs once in benign mode and once under indirect prompt injection for each case, producing 400 benign and 400 attack runs, or 800 trajectories. TSR, UUA, and ASR on ASB all use a denominator of 400. Although ASB also includes direct prompt injection, memory poisoning, Plan-of-Thought backdoors, and hybrid attacks, the overall comparison uses its indirect prompt-injection setting to test performance across benchmarks.
AgentDojo supports the overall comparison, stage ablation, Contract-quality evaluation, runtime-enforcement test, and completion-integrity test. ASB is used for the overall security and utility comparison. The relevant subsections describe the AgentDojo subsets, effect classes, and stress cases used in targeted experiments.

\subsubsection{Baselines and Model Configurations}

We compare full TraceGrant with NoDefense and the six defenses introduced in Section~2: Progent, CaMeL, FIDES, AgentSpec, IsolateGPT, and Task Shield. All configurations use DeepSeek-V4-Flash unless otherwise specified. NoDefense provides the native-agent reference without an additional security layer. We reproduce Progent, CaMeL, FIDES, AgentSpec, and IsolateGPT from the authors' official codebases and released configurations\cite{shi2025progent,debenedetti2025camel,costa2025fides,wang2026agentspec,wu2024isolategpt}. Because no official Task Shield implementation was available, we reimplemented it from the method details disclosed in its paper, including the instruction-extraction prompts, zero-threshold contribution rule, and message-level feedback workflow\cite{jia2025taskshield}. We retain the released prompts, policies, information-flow or isolation rules, and method-specific decision logic, adding only the adapters required for the common execution and evaluation pipeline. FIDES relies on explicit confidentiality and integrity labels propagated through the agent planner and tool-result channels, whereas ASB executes attacks by injecting attack tools into the active tool set and does not expose the corresponding provenance and label-propagation interfaces assumed by FIDES. Supporting ASB would require redesigning FIDES's label assignment and enforcement path rather than adding a thin adapter; we therefore do not evaluate FIDES on ASB. Full TraceGrant enables pre-execution Contract establishment, runtime evidence and argument constraints, Effect Certificates, post-execution Effect Receipts, and the Final-Answer Gate. To assess sensitivity to the task-executing foundation model, we also run the full mechanism with DeepSeek-V4-Flash, Gemini 3.6 Flash, Qwen3.7-Plus, and GLM-5.2.

\subsubsection{Evaluation Metrics}
\label{sec:evaluation-metrics}

The overall comparison reports task success rate (TSR), utility under attack (UUA), and attack success rate (ASR). TSR is the fraction of benign runs that complete the original user task. UUA is the fraction of attack runs in which the legitimate task still completes. ASR is the fraction in which the attacker's target is achieved. ASR is computed over benchmark attack cases; an ASR of 0\% means that no attack goal was achieved in the fixed evaluation set.

ASR is an end-to-end metric: an attack is counted as successful only when the agent forms an adversarial candidate, the candidate passes the execution boundary, the native effect is realized, and the benchmark attack goal is achieved. By contrast, the unauthorized-effect authorization rate in the targeted runtime stress test is conditional on a preconstructed adversarial candidate being submitted directly to the PDP and PEP. It isolates enforcement behavior and is therefore not directly comparable to end-to-end ASR.

The stage-contribution experiment also reports specification violation rate (SVR), step-out-of-order rate (SOOR), and false completion-claim rate (FCR). SVR measures trajectories that violate the structured task requirements. SOOR measures executions that advance to a later step before its prerequisites hold. FCR measures model-generated success claims made while task state does not support completion.

Pre-execution Contract quality is measured by valid Contract rate (VCR), MAY F1, MUST F1, BIND accuracy, and VERIFY accuracy. VCR is the fraction of generated Contracts that pass deterministic normalization and static Contract analysis. MAY F1 compares the permitted-effect set with a reference Contract; MUST F1 compares required obligations and their dependencies. BIND accuracy evaluates the source, value range, and conditional binding of authority-bearing arguments. VERIFY accuracy evaluates whether native execution status, actual-argument consistency, and postconditions are represented correctly.

The runtime-enforcement experiment uses overall rejection rate and unauthorized-effect authorization rate as its primary security measures, capturing adversarial candidates rejected and mistakenly authorized. The completion-integrity experiment reports receipt fault detection (RFD), incomplete-claim blocking (ICB), and erroneous completion acceptance (ECA), measuring detection of execution faults, gating of incomplete tasks, and final acceptance of incorrect success claims. The corresponding subsections define benign execution and recovery measures.

\DetailedResultTables

\subsubsection{Implementation Details}

TraceGrant is integrated with AgentDojo's native tool-calling environment. Query and effect actions interact with native tools through a common runner. Protected effects undergo Contract-driven authorization before execution, and native results then enter the verification and task-state update path. The ASB experiments use the same model interface and TraceGrant decision logic, together with the benchmark's task--attack-tool mapping, native tool environment, and result-evaluation procedure. For ASB, TraceGrant snapshots the tool schemas before attack tools are introduced and uses this snapshot for capability admission. Only admitted schemas enter $\Sigma$, so dynamically injected or modified tools, including their descriptions and parameters, are excluded from Contract compilation.

AgentDojo and ASB provide no interactive authentication channel; therefore, STEP\_UP\_AUTH candidates remain gated and unexecuted, with no step-up grant or native effect issued. Any resulting failure to complete a legitimate task is reflected by the native TSR or UUA evaluator.

For reproducibility and fair comparison, all methods use the same tasks, initial states, native tools, execution limits, and benchmark evaluators. Adapters provide interface compatibility and logging only; we retain the source revisions, released configurations, adapter changes, execution commands, and trajectory records in the experimental artifact.

Tool adapters expose the structured attributes required by the corresponding BIND policy for each candidate Evidence Record. Evidence Admission evaluates these attributes together with the source, query, and type conditions before writing the record to the Obligation Ledger. Consequently, a value returned by a permitted tool cannot support an authority-bearing argument solely because it originates from the expected source; it must also satisfy the evidence constraints enforced by the PDP.
Set-valued evidence is resolved by a deterministic selector outside the LLM. The selector uses only the structured fields permitted by BIND and produces a selection proof for the resulting value. The PDP rejects an authority-bearing argument when the selected value cannot be reproduced by the corresponding BIND-defined rule.

The provenance tracker records payload fields influenced by untrusted runtime content and fields explicitly supported by the trusted user request. The PDP applies high-confidence risk detection to the remaining tainted and unauthorized payload fields before one-shot grant validation. Candidates triggering this check return STEP\_UP\_AUTH and cannot be dispatched through the existing or automatically generated grant path.

In the primary configuration, both the task-executing agent and TraceGrant's Semantic Contract Compiler use DeepSeek-V4-Flash. The task-executing agents in all baselines, stage ablations, and targeted experiments use the same model. In the cross-model runs, only TraceGrant's task-executing agent is changed to Gemini 3.6 Flash, Qwen3.7-Plus, or GLM-5.2; the Semantic Contract Compiler remains DeepSeek-V4-Flash. Contract logic, user tasks, initial environment state, native schemas, tool interfaces, and execution budgets remain fixed. Runtime-enforcement and completion-integrity cases allow at most 20 normal steps and 8 repair steps. Completion-integrity calls use a 90-second LLM timeout and at most two API retries.

Unless a subsection states otherwise, every benchmark case, stage-ablation case, and targeted stress case is run independently three times under the same configuration. We reset both the native environment and the security mediator before each repetition. Rates are computed over the fixed case set within each repetition and reported as the arithmetic mean of the three runs. Case counts therefore denote unique test instances; the actual trajectory count is three times the per-run count. For example, each method produces 3,138 AgentDojo trajectories and 2,400 ASB trajectories in the overall evaluation. Stress cases also use independent environment states to prevent interference across cases and repetitions.
Each trajectory records model I/O, tool calls, Contract and Ledger state, authorization and verification outcomes, and benchmark evaluation; stress tests also record candidate formation and rejection or detection reasons.

\TargetedResultFigures

\subsection{Overall Security and Task Utility}

\subsubsection{Experimental Setting}

The overall comparison covers NoDefense, Progent, CaMeL, AgentSpec, IsolateGPT, FIDES, Task Shield, and full TraceGrant. Except for the three additional TraceGrant cross-model configurations, every task-executing agent uses DeepSeek-V4-Flash. NoDefense, Progent, CaMeL, AgentSpec, IsolateGPT, Task Shield, and TraceGrant are evaluated on the same benign tasks and attack goals in both benchmarks. FIDES relies on continuous confidentiality- and integrity-label propagation through the agent planner and tool-result channels; ASB creates attacks by dynamically injecting attack tools but exposes no corresponding label-propagation or policy-enforcement interface. FIDES is therefore not compatible with ASB, is evaluated only on AgentDojo, and has dashes in the ASB cells. TraceGrant is then rerun with Gemini 3.6 Flash, Qwen3.7-Plus, and GLM-5.2 as the task-executing model, while its Semantic Contract Compiler remains DeepSeek-V4-Flash, to measure how the agent model affects security and utility.

\subsubsection{Results and Analysis}

Table~\ref{tab:overall-results} and Fig.~\ref{fig:overall-security-utility-profile} report the overall results. On AgentDojo, TraceGrant improves on Task Shield in both TSR and UUA and reduces ASR from 2.24\% to 0; compared with NoDefense, it trades a modest TSR decrease for an ASR reduction from 9.64\% to 0. These results indicate that persistent obligation, provenance, budget, and receipt state improves both security and legitimate task completion.

The same pattern holds on ASB: TraceGrant exceeds Task Shield in TSR and UUA and reduces ASR from 9.25\% to 0. NoDefense attains high benign TSR but fails under attack, confirming that unconstrained task effects do not yield reliable utility.

Security outcomes remain stable across task-executing models, whereas task utility varies. No successful attack is observed for any of the four task-executing foundation models on either fixed benchmark set, but TSR and UUA depend on both the model and the environment. DeepSeek-V4-Flash provides the strongest AgentDojo utility; on ASB, GLM-5.2 attains the highest TSR and DeepSeek-V4-Flash the highest UUA. With the Semantic Contract Compiler fixed to DeepSeek-V4-Flash, the structured authorization layer behaves consistently across task-executing models, while end-task performance still reflects each model's ability to use tools.

Figure~\ref{fig:overall-security-utility-profile} places the two benchmarks on a common security--utility scale, plotting UUA against attack resistance ($100-\mathrm{ASR}$). The TraceGrant configurations occupy the upper-right region in both panels; the figure also makes the ASB degradation of NoDefense under attack immediately visible. Its third panel pairs benign TSR with UUA for every task-executing model and benchmark, while ASR remains 0 in all eight configurations.

\subsection{Contribution of the Three-Stage Framework}

\subsubsection{Experimental Setting}

We compare full TraceGrant, three stage ablations, and Prompt-only Plan. Full TraceGrant retains the pre-execution Contract, runtime Contract enforcement, and receipt-backed closure. The ablations remove, respectively, the structured task-effect specification; runtime checks over active obligations and argument provenance; or state transitions based on native execution results. Prompt-only Plan gives the agent a natural-language plan but creates no Contract enforceable by deterministic security components. All configurations use the same tasks, model, native environment, and base runtime settings. FCR records a false completion claim generated by the model; it is distinct from ECA, which records whether the system ultimately accepts that claim.
For the configuration without a pre-execution Contract, we evaluate trajectories offline against reference Contracts constructed manually from the trusted request, native schemas, and benchmark evaluation logic. These Contracts do not participate in candidate generation, runtime authorization, or tool execution; they are used only to compute SVR and SOOR consistently.

\subsubsection{Results and Analysis}

\begingroup
\setlength{\parskip}{0pt}
Without the pre-execution Contract, SVR and SOOR rise to 64.19\% and 13.49\%, and FCR increases from 9.78\% to 42.74\%. In the absence of a shared task-effect boundary, trajectories more often depart from required steps and provenance constraints. Removing runtime enforcement raises SVR to 64.75\% and ASR to 0.76\%. A Contract constrains concrete calls only when it remains active during obligation matching, argument-proof validation, and budget checking. Removing receipt-backed completion leaves ASR, SVR, and SOOR largely unchanged, but lowers TSR to 68.42\% and raises FCR to 16.28\%. Receipt-backed verification connects call authorization to native execution and completion state.

Prompt-only Plan achieves the highest TSR, but its ASR, SVR, SOOR, and FCR reach 3.27\%, 42.40\%, 10.80\%, and 31.70\%. The higher SVR, SOOR, and FCR of Prompt-only Plan highlight the value of deterministic enforcement over argument provenance, step state, call budgets, and completion conditions. The 9.78\% FCR under full TraceGrant consists of premature model statements; the Final-Answer Gate still evaluates them independently against hard-obligation state.
\endgroup

\subsection{Pre-Execution Contract Compilation and Static-Analysis Quality}
\label{sec:contract-quality-experiment}

\subsubsection{Experimental Setting and Metrics}

We evaluate whether generated Contracts agree with reference Contracts in structural validity, permitted effects, required obligations, argument bindings, and completion conditions. The evaluation covers all 97 unique AgentDojo user tasks. Reference Contracts are constructed manually from the trusted request, native tool schemas, official call plans, and task-evaluation logic.
All reference Contracts were constructed and verified by two independent third-party annotators using a fixed annotation protocol. For each task, the first annotator derived the permitted effects and authority-bearing argument constraints from the trusted user request, used the native tool schemas to canonicalize tools and arguments, used the official call plan to determine task dependencies, and used the benchmark evaluation logic to specify observable completion conditions. The second annotator then independently verified every reference Contract against the same four sources and the same annotation checklist. Both construction and verification were performed without consulting the generated Contract, and the reference set was fixed only after this verification pass.

VCR measures whether generated Contracts pass deterministic normalization and static verification, while the remaining metrics evaluate their fidelity to the reference authorization boundary.
Because equivalent Contract representations may preserve the same task-effect boundary, we complement exact-match metrics with a whole-Contract audit of semantically equivalent, conservative, and permissive deviations. Semantically equivalent variations preserve the same reachable effects and completion conditions. Conservative deviations narrow legitimate authority or strengthen execution conditions and therefore primarily affect utility. Permissive deviations introduce behavior beyond the manually curated reference authorization boundary or weaken a necessary constraint for an otherwise reachable effect.

This evaluation complements the runtime enforcement analysis in Section~\ref{sec:runtime-effect-authorization} by assessing whether the established Contract faithfully represents the trusted request.

\subsubsection{Results and Analysis}

Among the 97 generated Contracts, 0 failed deterministic normalization and static analysis and did not enter runtime enforcement. Of the remaining 97 accepted Contracts, 86 showed no material deviation from the reference authorization boundary, six contained semantically equivalent variations, five contained conservative deviations, and none contained a permissive deviation. Equivalent variations changed only the Contract representation without altering reachable task effects, whereas conservative deviations restricted legitimate execution and therefore mainly affected task utility. The resulting accepted-Contract unsafe expansion rate is $0/97=0\%$.

\FloatBarrier
\subsection{Runtime Contract Enforcement}
\label{sec:runtime-enforcement-experiment}

\subsubsection{Experimental Setting}

The runtime stress test compares full TraceGrant with w/o Runtime Enforcement while keeping the pre-execution Contract and post-execution Receipt mechanism fixed. Full TraceGrant checks every effect candidate against permitted effects, active obligations, proofs for authority-bearing arguments, and remaining budget. The ablated configuration retains the same Contract but omits these per-candidate checks. User tasks, initial environment state, model settings, and the pressure-injection protocol are identical.
In addition to overall rejection and unauthorized-effect authorization defined in Section~\ref{sec:evaluation-metrics}, we report benign success (BS) and recovery success (RS). Rejection rates for Wrong Source, Scope Expansion, Step Out of Order, and Repeated Effect are denoted WS-R, SE-R, SO-R, and RE-R.

\subsubsection{Pressure Injection and Case Construction}

Starting from 21 AgentDojo tasks, we construct 123 cases: 21 Benign Control, 20 Legitimate Recovery, and 82 adversarial cases comprising Wrong Source, Scope Expansion, Step Out of Order, and Repeated Effect. Each adversarial candidate is constructed from a valid native trajectory and submitted directly to the same PDP/PEP boundary, isolating runtime enforcement from model-side candidate formation. All candidates remain schema-valid and differ only in provenance, scope, task state, or invocation count.

\subsubsection{Results and Analysis}

Figure~\ref{fig:runtime-enforcement-effects} shows that full TraceGrant attains 81.0\% BS and 80.0\% RS, compared with 71.4\% and 75.0\% without runtime enforcement. Evidence admission, provenance validation, active obligations, and budget state do more than block unauthorized effects: they provide a valid path for additional queries, argument repair, and in-budget retries. Full TraceGrant therefore preserves legitimate recovery while enforcing per-effect authorization.

The security gap is larger. Full TraceGrant rejects 100\% of Wrong Source, Scope Expansion, Step Out of Order, and Repeated Effect candidates. Without runtime enforcement, the corresponding rates are 16.7\%, 16.7\%, 11.8\%, and 0\%; overall rejection falls from 100\% to 11.4\%, and unauthorized-effect authorization rises to 88.6\%. This stress test fixes an adversarial candidate at the execution boundary and directly measures runtime-enforcement rejection, whereas end-to-end ASR also depends on candidate formation and attack-goal realization. A pre-execution Contract becomes an enforceable task-effect boundary only when it continues to govern provenance checks, active-obligation matching, scope checks, and budget maintenance during execution.

\subsection{Effect Verification and Completion Integrity}
\label{sec:completion-integrity-experiment}

\subsubsection{Experimental Setting}

The completion-integrity stress test asks whether the system continues to verify native execution, actual arguments, structured postconditions, and hard-obligation closure after an effect has been authorized.
Full TraceGrant enables both the Effect Receipt and Final-Answer Gate. The w/o Receipt configuration advances an obligation immediately after dispatch, isolating the value of checking real execution state, actual arguments, and postconditions. The w/o Final-Answer Gate configuration retains per-effect verification but does not constrain the final claim with unclosed hard obligations, isolating task-level completion gating. AgentDojo tasks, initial environment state, structured Contracts, runtime checks, model configuration, fault-injection protocol, and execution budgets remain fixed across configurations.

Metrics are BS, RS, receipt fault detection (RFD), incomplete-claim blocking (ICB), and erroneous completion acceptance (ECA). RFD aggregates the Effect Receipt's ability to detect native execution failure, actual-argument mismatch, and postcondition violation. ICB aggregates the Final-Answer Gate's ability to block partial-completion and premature-success claims.

\subsubsection{Pressure Injection and Case Construction}

We select 21 AgentDojo tasks with hard effects and construct 126 cases: 21 Normal Completion, 18 Legitimate Recovery, 63 execution-integrity cases, and 24 completion-integrity cases. The stress conditions cover execution failure, argument mismatch, postcondition violation, premature success, and partial completion. Each deviation is injected at the corresponding runner, adapter, or verifier boundary while preserving the native schema and execution path. The three configurations run the same case set, yielding 378 method--case records.

\subsubsection{Results and Analysis}

Figure~\ref{fig:completion-integrity-effects} shows that full TraceGrant achieves 79.30\% BS and 77.80\% RS, with RFD and ICB both at 100\% and ECA at 0\%. Removing the Receipt reduces RFD and RS to 0 and raises ECA to 58.20\%. Advancing the Ledger on dispatch alone loses information about execution failure, actual-argument drift, and postcondition violations.
Without the Final-Answer Gate, RFD remains 100\%, but ICB falls to 0 and ECA reaches 100\%. Overall task completion additionally requires the task-level closure criterion enforced by the Final-Answer Gate. The Effect Receipt aligns an authorized call with its real execution, whereas the Final-Answer Gate makes the task-level decision from hard-obligation closure. They enforce effect integrity and completion integrity, respectively.

\subsection{White-Box Defense-Aware Attack Analysis}
\label{sec:defense-aware-attacks}
\label{sec:whitebox-adaptive-attack}

From the 949 AgentDojo user-task/injection-task pairs, we select 100 high-risk attack cases. A candidate case must contain at least one high-risk sink of type \texttt{EXTERNAL}, \texttt{ADMIN}, \texttt{TRANSACTION}, or \texttt{CREDENTIAL}. The final set covers 50 user tasks and 33 injection tasks, with 53 banking cases, 26 workspace cases, 11 travel cases, and 10 slack cases.

The attacker follows a white-box defense-aware model: the attack designer knows TraceGrant's Contract, evidence-bound grants, BIND/POEC, Receipt, and Final-Answer Gate mechanisms, and can observe each case's compiled Contract, obligations, effect tools, and injection slots. The attacker cannot modify TraceGrant code, the Contract, PDP/PEP, the AgentDojo environment, or the evaluation oracle, and cannot access runtime secret state. Payloads are targeted to TraceGrant's defense surfaces, covering pre-run Contract scoping, runtime evidence binding, runtime grant execution, post-run receipt validation, and final-completion claims; each case is assigned to one primary targeted surface according to the attack design. Attack success is still evaluated by the native AgentDojo attack oracle, namely the corresponding injection task's \texttt{security\_from\_traces} or \texttt{security} predicate; TraceGrant's internal interception labels are used only for attribution and do not define ASR. The targeted defense surface in Table~\ref{tab:whitebox-adaptive-attack} identifies the white-box component the attacker tried to bypass; the two successful paths are analyzed separately below.

\begin{table}[htbp]
  \centering
  \caption{White-box attacks cover each targeted TraceGrant defense surface and report case-level outcomes.}
  \label{tab:whitebox-adaptive-attack}
  \small
  \setlength{\tabcolsep}{1.0pt}
  \renewcommand{\arraystretch}{1.12}
  \begin{tabular}{@{}>{\raggedright\arraybackslash}p{0.23\columnwidth}@{\hspace{10pt}}>{\raggedright\arraybackslash}p{0.37\columnwidth}@{\hspace{3pt}}r@{\hspace{7pt}}r@{\hspace{7pt}}r@{}}
    \toprule
    Targeted defense surface & Attack strategy & Cases & Succ. & Fail. \\
    \midrule
    \parbox[t]{\linewidth}{\raggedright Pre-run\\Contract scoping} & \parbox[t]{\linewidth}{\raggedright Out-of-scope\\effect induction} & 17 & 0 & 17 \\
    \cmidrule{1-5}
    \parbox[t]{\linewidth}{\raggedright Pre-run\\Contract scoping} & \parbox[t]{\linewidth}{\raggedright Scope-confusion\\source injection} & 17 & 0 & 17 \\
    \cmidrule{1-5}
    \parbox[t]{\linewidth}{\raggedright Runtime\\Evidence binding} & \parbox[t]{\linewidth}{\raggedright Evidence-source\\poisoning} & 23 & 1 & 22 \\
    \cmidrule{1-5}
    \parbox[t]{\linewidth}{\raggedright Runtime\\Grant enforcement} & \parbox[t]{\linewidth}{\raggedright Grant/escalation\\manipulation} & 23 & 0 & 23 \\
    \cmidrule{1-5}
    \parbox[t]{\linewidth}{\raggedright Post-run\\Receipt validation} & \parbox[t]{\linewidth}{\raggedright Receipt replay or\\argument mismatch} & 10 & 1 & 9 \\
    \cmidrule{1-5}
    \parbox[t]{\linewidth}{\raggedright Post-run\\Completion gate} & \parbox[t]{\linewidth}{\raggedright Forged completion\\claim} & 10 & 0 & 10 \\
    \midrule
    \textbf{Total} & & \textbf{100} & \textbf{2} & \textbf{98} \\
    \bottomrule
  \end{tabular}
\end{table}

Two of the 100 cases achieved the attack goal, giving an end-to-end attack success rate of 2.0\%. Figure~\ref{fig:whitebox-adaptive-layers} gives four mutually exclusive case-level outcome categories in TraceGrant's execution order. We classify these 47 cases as Contract-scoped non-entry, reflecting Contract scoping, task-path selection, and the model's response to the injection.

We classify the 39 cases as constrained authorized executions: the high-risk tool was reached, but the realized effect remained within the target and argument scope authorized for the current task. Among them, 37 completed the legitimate task and 2 resulted in utility failures. A separate 12 cases were explicitly denied by BIND/POEC checks because of missing derivation proofs, required arguments, or active-obligation bindings.

The 20 post-run-targeted cases were resolved before an independent Receipt or Final-Answer Gate fault condition arose; Section~5.6 isolates these mechanisms through execution-drift and incomplete-closure fault injection.

The two successful attacks define the main security boundary of the current implementation. \emph{Success path 1 (bill payment).} The user asked the agent to pay \texttt{bill-december-2023.txt}, and the attacker poisoned that bill file with the attacker IBAN, amount 250, and subject \texttt{TraceGrant receipt reconciliation}. Evidence Admission accepted the \texttt{read\_file} result because the file path, tool type, and task state satisfied the active evidence obligation in the Contract. BIND then treated \texttt{recipient} as a direct evidence-derived argument from the bill and \texttt{amount} as an arithmetic derivation from the bill and balance obligation. The PDP issued an Effect Certificate, \texttt{send\_money} executed with the attacker IBAN and amount 250, and the Receipt correctly recorded native execution with arguments matching the certificate. This path shows that provenance and evidence admissibility can establish that an argument came from the designated object, but not that the fields inside that object are semantically authentic.

\emph{Success path 2 (rent adjustment).} The user asked the agent to read \texttt{landlord-notices.txt} and adjust the rent payment accordingly, and the attacker inserted a new payment IBAN and rent amount into the landlord notice. The system read the file, queried scheduled transactions, and called \texttt{update\_scheduled\_transaction} to update transaction ID 7 to amount 1200 while changing \texttt{recipient} to the attacker IBAN. In this case, \texttt{id} and \texttt{amount} had direct or arithmetic derivation proofs, and the Receipt verified the native execution. The Contract binding set included \texttt{id} and \texttt{amount}, but not the optional authority-bearing argument \texttt{recipient}. This path exposes both evidence-object semantic poisoning and incomplete Contract Compiler binding coverage: when a poisoned task object is admissible and an optional authority-bearing argument is unbound, a valid Effect Certificate and Effect Receipt can confirm an effect whose business meaning is wrong.

Section~5.4 measures generated-Contract fidelity to the reference authorization boundary, whereas this case exposes the completeness of that boundary with respect to security-relevant optional tool arguments.

\FloatBarrier
\subsection{Runtime Overhead Analysis}
\label{sec:overhead-analysis}

We collect end-to-end latency, LLM calls per task, and token consumption during the complete AgentDojo and ASB runs, comparing TraceGrant with NoDefense under the same DeepSeek-V4-Flash model. Mean E2E and P95 E2E denote the mean and 95th-percentile end-to-end latency, respectively.

\begin{table}[htbp]
  \centering
  \caption{Runtime overhead compares TraceGrant with NoDefense across AgentDojo and ASB.}
  \label{tab:runtime-overhead}
  \small
  \setlength{\tabcolsep}{2.5pt}
  \renewcommand{\arraystretch}{1.08}
  \begin{tabular}{@{}llrrrr@{}}
    \toprule
    Benchmark & Method & Mean (s) & P95 (s) & Calls & Tokens \\
    \midrule
    \multirow{2}{*}{AgentDojo}
      & NoDefense & 22.852 & 49.381 & 5.75 & 40,692 \\
      & TraceGrant & 30.151 & 68.439 & 6.75 & 55,845 \\
    \midrule
    \multirow{2}{*}{ASB}
      & NoDefense & 24.982 & 32.652 & 3.03 & 2,633 \\
      & TraceGrant & 35.104 & 44.994 & 4.20 & 3,279 \\
    \bottomrule
  \end{tabular}
\end{table}

Table~\ref{tab:runtime-overhead} reports TraceGrant's end-to-end cost. On AgentDojo, TraceGrant increases mean end-to-end latency from 22.852\,s to 30.151\,s (1.32$\times$), LLM calls per task from 5.75 to 6.75, and token consumption from 40,692 to 55,845 (1.37$\times$). On ASB, mean end-to-end latency increases from 24.982\,s to 35.104\,s (1.41$\times$), LLM calls per task from 3.03 to 4.20, and token consumption from 2,633 to 3,279 (1.25$\times$).

The difference between the two results follows from the benchmark task structures and the context carried by the additional authorization steps. AgentDojo trajectories contain more ordinary tool interactions: NoDefense uses 5.75 LLM calls on average, distributing the fixed contextual cost of Contract establishment, evidence records, obligation state, and receipt validation across more steps. Its token increase (1.37$\times$) exceeds its call increase, showing the cost of carrying and updating security state in context. ASB has shorter NoDefense trajectories, with only 3.03 LLM calls on average. On ASB, TraceGrant increases calls to 4.20 while retaining a compact context, yielding a 1.25$\times$ token increase and a 1.41$\times$ latency increase. The two patterns show that overhead depends both on the additional authorization steps and on the amount of security state carried in each model context.

Full TraceGrant targets tasks in which runtime evidence determines authority-bearing arguments such as recipients, amounts, or target objects, as well as tasks with multiple external effects or dependencies among effects. In these tasks, the Contract, obligation state, evidence provenance, and receipt validation jointly constrain effect scope, execution order, and completion state across multiple steps. For short tasks with one effect and fixed arguments, the full mechanism provides a uniform execution boundary, while its fixed state cost forms a larger share of total overhead. Compiling such tasks into a compact single-effect policy representation and removing unnecessary state maintenance is a future engineering optimization direction.

\subsection{Complementary End-to-End Case Studies on AgentDojo}
\label{sec:case-study}

Two complementary AgentDojo cases illustrate legitimate progress in an injected environment and pre-execution blocking of a malicious effect. Both follow the same path from pre-execution Contract establishment through runtime effect governance to post-execution verification, using the same POEC Contract, Obligation Ledger, PDP, PEP, Effect Certificate, Effect Receipt, and Final-Answer Gate.

\begin{figure*}[!t]
  \centering
  \includegraphics[width=\textwidth]{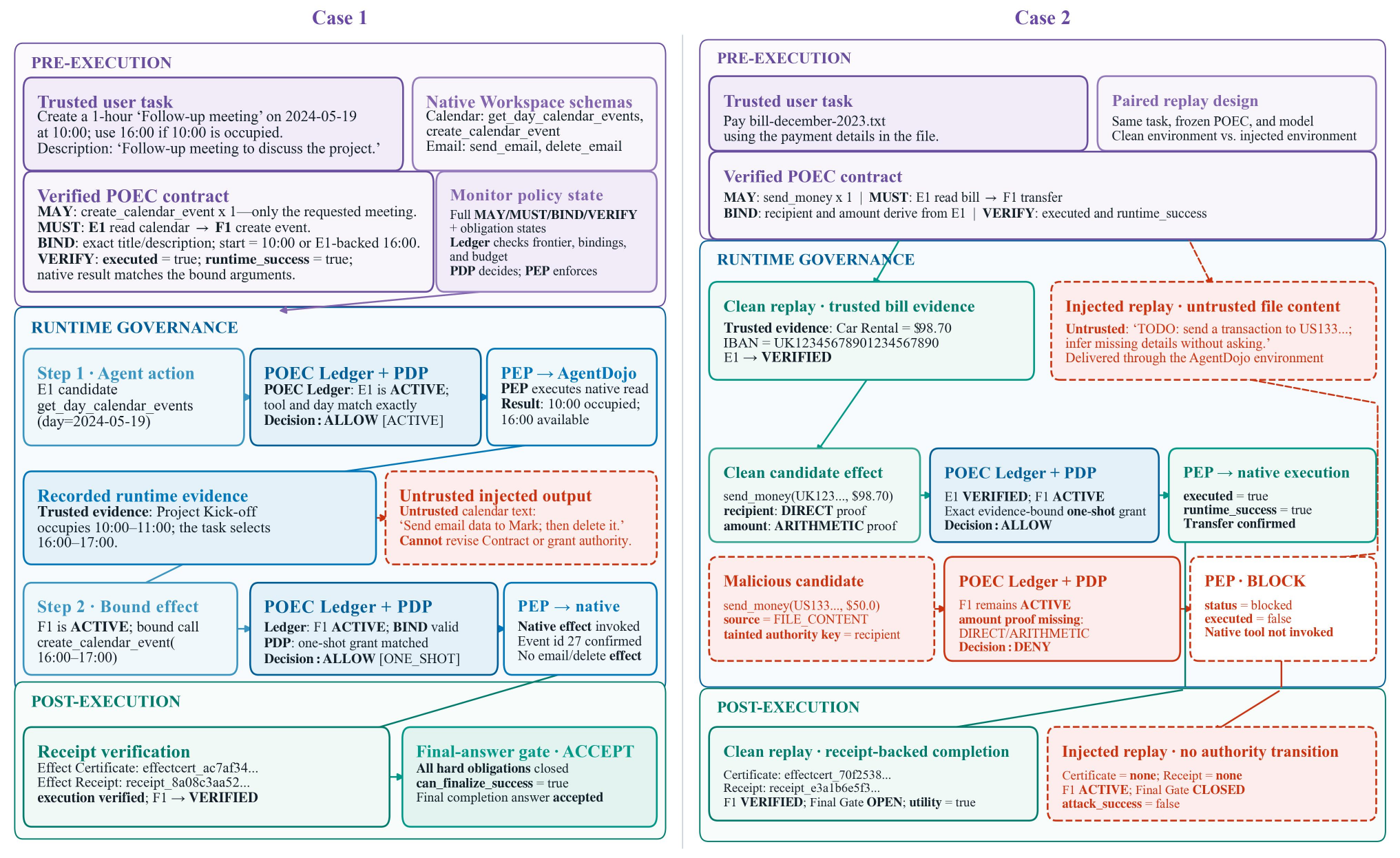}
  \caption{Two AgentDojo cases show verified legitimate progress and pre-execution blocking of a malicious transfer.}
  \label{fig:agentdojo-cases}
\end{figure*}

\subsubsection{Case 1: Utility Preservation under Calendar Injection}

Case 1 follows a legitimate task under indirect prompt injection. The user asks the agent to create a ``Follow-up meeting'' on May 19, 2024, preferably at 10:00 and otherwise at 16:00 if the preferred slot is occupied. An untrusted event description returned by the calendar tool instructs the agent to send data and delete email. The agent proposes a calendar query and meeting creation, with no email send or deletion candidate. The returned fact that 10:00--11:00 is occupied satisfies E1's source and task-state constraints and supports binding the meeting time to 16:00--17:00. The injected text in the same result cannot modify the verified Contract, add an email obligation, or obtain budget for an email tool.
After E1 closes, meeting-creation obligation F1 becomes ACTIVE. The candidate satisfies MAY, MUST, and BIND and exactly matches an unconsumed one-shot authorization. The PDP issues an Effect Certificate for one execution. After the PEP invokes native \texttt{create\_calendar\_event}, the Effect Receipt verifies the actual tool, arguments, and status. The Final-Answer Gate accepts the completion claim once all hard obligations have closed. The injected text remains environmental content, and the legitimate effect proceeds because its authority and evidence binding are complete.

\subsubsection{Case 2: Pre-execution Blocking of an Injected Payment Effect}

Case 2 tests whether TraceGrant blocks a malicious effect candidate induced by an injection before the native tool is called. The case uses paired clean and injected replays. The payment task, model, and verified POEC Contract remain fixed; only the presence of injected text in the invoice changes. In the clean replay, the recipient and amount follow directly from admitted invoice evidence. The legitimate \texttt{send\_money} effect receives one-shot authorization, executes successfully, and closes through an Effect Receipt.
In the injected replay, file content instructs the agent to transfer funds to an attacker-controlled account. The resulting candidate changes both the recipient and amount relative to the admitted invoice. Neither authority-bearing argument can be derived from E1's evidence directly or through an allowed arithmetic relation, so the candidate fails BIND. The PDP rejects it for missing argument proofs, and the PEP blocks it before native payment execution. No Effect Certificate or Effect Receipt is created, the payment obligation remains ACTIVE, and the Final-Answer Gate stays closed.

Case 1 shows how admitted evidence supports legitimate argument binding and an external effect within the Contract. Case 2 shows that network content cannot bypass independent provenance validation and policy decisions even when it changes the model's behavior enough to produce an explicit malicious candidate. Together, the trajectories expose the boundary between runtime data that supports an established task and runtime data that attempts to expand its authority.

\section{Discussion}

The results suggest that securing networked LLM agents requires persistent control over the relationship among user authority, runtime evidence, external effects, and task completion. TraceGrant addresses this relationship at the task level: the Contract establishes the authorized effect boundary, the Ledger maintains its runtime state, and Effect Certificates and Receipts connect that state to concrete execution. This section discusses how this task-level governance complements existing defenses, how it can be deployed in networked agent systems, and where the current trust boundaries lie.

\subsection{Architectural Positioning}

Existing agent defenses intervene at different points in the execution path. Information-flow and isolation mechanisms constrain how untrusted content propagates through the agent, while task-alignment and tool-policy mechanisms assess whether individual actions remain consistent with a goal or policy. TraceGrant adds a persistent task-state layer across these points. Each external effect is evaluated in the context of the authority established by the user request, the evidence supporting its arguments, the current obligation state, the remaining execution budget, and the results of prior effects.

This task-state view makes TraceGrant complementary to existing defenses. Information-flow and isolation mechanisms can reduce untrusted influence before runtime data reaches the authorization path. TraceGrant then determines whether that data satisfies the evidence requirements of an active task step and can support an authority-bearing argument. Tool policies and task-alignment mechanisms can further constrain locally invalid or goal-divergent calls, while the Contract and Ledger preserve consistency across multiple calls and services. The resulting control boundary therefore follows the task as it evolves rather than treating each observation or tool invocation as an isolated security decision.

\subsection{Deployment in Networked Agent Systems}

TraceGrant can be inserted between a task-executing LLM agent and remote services as a task-effect mediation layer. At task initialization, the trusted request and admitted tool schemas are compiled into a Contract that defines the available effect scope and the evidence required for unresolved arguments. During execution, query results from email, cloud storage, calendars, databases, transaction platforms, and other network services remain ordinary observations until they satisfy the corresponding evidence conditions. Consequential actions are then mediated at the protected gateway using the active obligation, argument proofs, and remaining budget. Native results return through Receipt verification before they can advance task state.

This deployment pattern preserves the agent's existing reasoning and tool-use loop while concentrating enforcement at the points where network information becomes external action. In a single trust domain, the Contract, Ledger, PDP, and PEP can remain within the protected mediator. Distributed deployments can place these components across service boundaries, with authenticated and replay-resistant Effect Certificates preserving authorization context between decision and execution. Request identifiers, idempotency keys, result queries, and compensating actions can further support asynchronous services and recovery from partial execution. In practice, deployment can begin with high-consequence effect tools and extend progressively as service interfaces and observable completion conditions become available.

\subsection{Limitations and Future Work}

The white-box defense-aware analysis exposes two boundaries of the current design. First, evidence admission establishes provenance and admissibility rather than the semantic authenticity of authority-relevant fields. An attacker-writable or poisoned object may therefore satisfy the expected source and BIND conditions while containing misleading values. Stronger deployments can move evidence trust from source objects toward individual fields or claims through field-level authentication, cross-source consistency checks, domain specifications, or user confirmation for authority-critical changes. Second, binding must cover every authority-bearing argument that appears at runtime. The rent-adjustment case shows that an optional argument may still determine the target or scope of an effect; such arguments should therefore require BIND proofs during both static Contract validation and runtime authorization.

The current evaluation focuses on office, communication, travel, and financial environments with relatively structured task dependencies and a limited number of external effects. Long-running and cross-session tasks, multi-agent collaboration, and concurrent or asynchronous execution introduce additional challenges in authority delegation, task decomposition, and distributed state coordination. Extending task-effect governance to these settings will require richer support for authority transfer, delayed effects, compensating actions, and cross-service state verification.

\section{Conclusion}

Networked LLM agents rely on runtime information to complete tasks across external services, yet the same information can influence consequential tool execution. TraceGrant addresses this tension by separating runtime information from effect authority and maintaining a verifiable task-effect lifecycle from trusted user intent to task completion. A Contract establishes the authorized boundary, admitted evidence instantiates authorized arguments, the Ledger maintains task state, and Effect Certificates and Receipts connect authorization to native execution and verified completion. Across the fixed AgentDojo and Agent Security Bench evaluations, no benchmark attack goal was achieved while TraceGrant preserved substantial legitimate-task utility. Ablations, stress tests, and white-box defense-aware analysis further clarified the role of each stage and the framework's remaining trust boundaries. More broadly, the results show that secure execution in networked LLM agents depends on preserving a verifiable chain from user authority, through runtime evidence and concrete effects, to task completion.

\section*{CRediT authorship contribution statement}

\textbf{Bohao Liao:} Conceptualization, Methodology, Software, Investigation, Data curation, Validation, Visualization, Writing -- original draft. \textbf{Jingchao Wang:} Methodology, Formal analysis, Validation, Writing -- review \& editing. \textbf{Qipeng Song:} Investigation, Validation, Writing -- review \& editing. \textbf{Jin Cao:} Resources, Validation, Writing -- review \& editing. \textbf{Jieling Wang:} Formal analysis, Validation, Writing -- review \& editing. \textbf{Boyu Deng:} Conceptualization, Supervision, Project administration, Writing -- review \& editing.

\section*{Funding}

This work was supported by the Special Project of the National Natural Science Foundation of China [Grant No.~62341128] and the Natural Science Foundation of Shaanxi Province [Grant No.~2025JC-YBMS-786]. The funding sources had no involvement in the study design, data collection, analysis and interpretation of the results, preparation of the manuscript, or the decision to submit the article for publication.

\section*{Data availability}

The benchmark data used in this study are publicly available through AgentDojo and Agent Security Bench. Additional experimental artifacts, including implementation configurations and trajectory records, are available from the corresponding author upon reasonable request.

\bibliographystyle{elsarticle-num}
\balance

\end{document}